\documentclass[11pt]{article}

\usepackage[paper=a4paper,dvips,top=2.3cm,left=2.3cm,right=2.3cm,bottom=2.9cm,footskip=1.6cm]{geometry}
\usepackage{amsfonts,amsmath,amsthm,amssymb}
\numberwithin{equation}{section}  
\usepackage[svgnames]{xcolor}
\usepackage{fix-cm}
\usepackage{sectsty}
\usepackage{fancyhdr}
\usepackage{lastpage}
\usepackage{graphicx}
\usepackage{url}
\usepackage{enumerate}
\usepackage{paralist}
\usepackage{multicol}
\usepackage{color}  
\usepackage{float}  
\usepackage{epsfig}
\usepackage{hyperref}   
\hypersetup{colorlinks=true,linkcolor=beamer@PRD, citecolor=beamer@PRD}
\usepackage{authblk}  
\usepackage{cite}  

\definecolor{beamer@blue}{RGB}{0,0,255}
\definecolor{beamer@mediumblue}{RGB}{0,0,190}
\definecolor{beamer@midnightblue}{RGB}{25,25,112}
\definecolor{beamer@navy}{RGB}{0,0,128}
\definecolor{beamer@darkblue}{RGB}{0,0,139}
\definecolor{beamer@purple}{RGB}{128,0,128}
\definecolor{beamer@levander}{RGB}{100.,149.,237.}
\definecolor{beamer@PRD}{RGB}{46,48,146}
\definecolor{beamer@green}{RGB}{0,128,0}
\definecolor{beamer@darkgreen}{RGB}{0,100,0}
\definecolor{beamer@olive}{RGB}{128,128,0}
\definecolor{beamer@darkolivegreen}{RGB}{85,107,47}
\definecolor{beamer@gray}{RGB}{190,190,190}
\definecolor{beamer@ivry}{RGB}{220,220,220}
\definecolor{beamer@new}{RGB}{40,120,50}
\definecolor{shadecolor}{RGB}{220,220,220}
\definecolor{beamer@darkslategray}{RGB}{47,79,79}
\definecolor{beamer@chocolate}{RGB}{210,105,30}
\definecolor{beamer@brown}{RGB}{165,42,42}
\definecolor{beamer@orangered}{RGB}{255,69,0}
\definecolor{beamer@maroon}{RGB}{128,0,0}
\definecolor{beamer@white}{RGB}{234,242,243}
\definecolor{beamer@silver}{RGB}{0.5,0.45,0.37}

\begin{document}


\title{\textbf{Higher order nonclassicality from nonlinear coherent states for models with quadratic spectrum}}


\author{\textbf{Anaelle Hertz$^1$, Sanjib Dey$^{2,3}$, V\'eronique Hussin$^{2,4}$ and Hichem Eleuch$^{5,6}$}}
\date{\footnotesize{$^{1}$Centre for Quantum Information and Communication, \'Ecole Polytechnique, \\ Universit\'e Libre de Bruxelles, Bruxelles 1050, Belgium \\ $^{2}$Centre de Recherches Math\'ematiques, Universit\'e de Montr\'eal, Montr\'eal H3C 3J7, Qu{\'e}bec, Canada \\ $^{3}$Department of Mathematics and Statistics, Concordia University, Montr\'eal H3G 1M8, Qu\'ebec, Canada \\ $^{4}$Department de Math\'ematiques et de Statistique, Universit\'e de Montr\'eal, Montr\'eal H3C 3J7, Qu\'ebec, Canada \\ $^{5}$Department of Physics, McGill University, Montr\'eal H3A 2T8, Qu\'ebec, Canada \\ $^{6}$Institute for Quantum Science and Engineering, Texas A$\&$M University, College Station, Texas 77843, USA \\ E-mail: ahertz@ulb.ac.be, dey@crm.umontreal.ca, veronique.hussin@umontreal.ca, heleuch@fulbrightmail.org}}
\maketitle
    	
\thispagestyle{fancy}
\begin{abstract}
Harmonic oscillator coherent states are well known to be the analogue of classical states. On the other hand, nonlinear and generalised coherent states may possess nonclassical properties. In this article, we study the nonclassical behaviour of nonlinear coherent states for generalised classes of models corresponding to the generalised ladder operators. A comparative analysis among them indicates that the models with quadratic spectrum are more nonclassical than the others. Our central result is further underpinned by the comparison of the degree of nonclassicality of squeezed states of the corresponding models.
\vspace{0.3cm}

\end{abstract}	 
\begin{section}{Introduction} \label{sec1}
\addtolength{\footskip}{-0.4cm} 
\addtolength{\voffset}{1.2cm} 
In 1926, Erwin Schr\"odinger first introduced coherent states while searching for classical like states \cite{Schrodinger}. In the very first proposal, coherent states were interpreted as nonspreading wavepackets when they move in the harmonic oscillator potential. They minimise the Heisenberg's uncertainty relation, with equal uncertainties in each quadrature. Thus, they are the best quantum mechanical representation of a point in phase space, or in other words, they are the closest possible quantum mechanical states whose behaviour resemble  that of classical particles. Coherent states were introduced later by Glauber and Sudarshan \cite{Glauber,Sudarshan2} as a starting point of nonclassicality in terms of the $P$-function for arbitrary density matrices $\rho$,
\begin{eqnarray}
\rho = \int P(z)\vert z\rangle\langle z\vert d\text{Re}z~d\text{Im}z, \quad \int P(z) d\text{Re}z~d\text{Im}z=1.
\end{eqnarray}
For coherent states, the weight function $P(z)$ can be interpreted as a probability density, as in this case the $P$-function is a delta function. Glauber defined the nonclassical states as those for which the $P$-distribution fails to be a probability density. More specifically, if the singularities of $P$-functions are either of types stronger than those of the delta functions or they are negative, the corresponding states have no classical analogue. A different argument is given in \cite{Johansen}, where the author says that a quantum state may be nonclassical even though the $P$-distribution is a probability density, rather, the nonclassicality is associated with the failure of the Margenau-Hill distribution to be a probability distribution. Nevertheless, we will stick to the convention of Glauber throughout our discussion. It is important to know that while the coherent states play important roles in various branches of physics due to their classical like behaviour, the nonclassical states also have several interesting characteristics for the purpose of quantum information processing. Over last few decades, there have been numerous experimental attempts in search of nonclassical states \cite{Meekhof_Monroe_King_Itano_Wineland,Bose_Jacobs_Knight}. The underlying inspiration behind all these efforts is that the nonclassical states are the prerequisites for creating entangled states \cite{Kim_Son_Buzek_Knight}, which are the most fundamental requirements for quantum teleportation.

There exist different types of \textit{nonclassical states} in the literature; such as, squeezed states \cite{Walls,Loudon_Knight}, photon added coherent states \cite{Agarwal_Tara}, Schr\"odinger cat states \cite{Xia_Guo}, pair coherent states \cite{Agarwal_Biswas}, photon subtracted squeezed states \cite{Wakui_Takahashi_Furusawa_Sasaki} and many more. For a concrete review on the subject, one can follow; for instance, \cite{Dodonov}. Apart from the aforementioned nonclassical states, the nonclassical properties of different type of coherent states have also been explored, for instance, for Gazeau-Klauder coherent states \cite{Roy_Roy_Morse}, nonlinear coherent states \cite{Roy_Roy_Nonlinear,Choquette_etal,Obada_etal,Recamier_etal,Tavassoly} and coherent states in noncommutative space \cite{Dey_Hussin,Dey_Fring_Hussin}. In this article, we explore the comparative analysis of nonclassical nature of nonlinear coherent states for two different type of models associated to the nonlinear ladder operators, and realise that the models with quadratic spectrum produce higher amount of nonclassicality than the others. 

Our manuscript is organised as follows: In Sec. \ref{sec2}, we discuss the general construction procedure of nonlinear coherent and squeezed states. In Sec. \ref{sec3}, we revisit a well known technique of computing quantum entanglement from nonclassical states by utilising a quantum beam splitter. In Sec. \ref{sec4}, nonclassical properties of nonlinear coherent states for two general classes of models are analysed. Our conclusions are stated in Sec. \ref{sec5}.
\end{section}
\begin{section}{Nonlinear coherent and squeezed states}\label{sec2}
We start with a brief review of coherent states $\vert z\rangle$, which are defined as the eigenstates of the boson annihilation operator $a\vert z\rangle=z\vert z\rangle$, with $z$ being complex eigenvalues \cite{Glauber}. In terms of Fock states they are represented in a compact form by
\begin{eqnarray}\label{coherent}
\vert z\rangle &=& \frac{1}{\mathcal{N}(z)}\displaystyle\sum_{n=0}^{\infty}\frac{z^n}{\sqrt{n}!}\vert n\rangle, \qquad \mathcal{N}(z)=e^{\frac{\vert z\vert^2}{2}}.
\end{eqnarray}
It was shown that the above expression is equivalent to the one obtained by applying the Glauber's unitary displacement operator, $D(z)=\exp(z a^\dagger-z^\ast a)$, on the vacuum $\vert 0\rangle$. The striking feature of coherent states is that they minimise the uncertainty relation, i.e. $\Delta x^2\Delta p^2=1/4$, with equal uncertainties in each coordinate and, therefore, exhibit reduced noise in optical communications \cite{Gerry_Knight_Book}. On the other hand, it is well known that for squeezed states the uncertainty in one of the coordinates $x,p$ becomes squeezed, so that they produce much lower noise in the corresponding quadrature in comparison to the coherent states \cite{Walls,Loudon_Knight}. Squeezed states $\vert z,\gamma\rangle$ are constructed in two equivalent ways as for coherent states. First, from the eigenvalue definition, $(a+\gamma a^\dagger)\vert z,\gamma\rangle = z\vert z,\gamma\rangle$ \cite {Fu_Sasaki,Alvarez_Hussin} and, second, by acting the squeezing operator, $S(\gamma)=\exp[(\gamma a^\dagger a^\dagger-\gamma^\ast aa)/2]$, on the coherent state, i.e. $\vert z,\gamma\rangle=S(\gamma)D(z)\vert 0\rangle$, with $ z,\gamma\in\mathbb{C},\vert\gamma\vert<1$. Consequently, one can express the squeezed states in the number state basis as follows
\lhead{Higher order nonclassicality from nonlinear coherent states}
\chead{}
\rhead{}
\begin{eqnarray}\label{squeezed}
\vert z, \gamma\rangle &=& \frac{1}{\mathcal{N}(z,\gamma)}\displaystyle\sum_{n=0}^{\infty}\frac{1}{\sqrt{n}!}\left(\frac{\gamma}{2}\right)^{n/2}\mathcal{H}_n\left(\frac{z}{\sqrt{2\gamma}}\right)\vert n\rangle,
\end{eqnarray}
where, $\mathcal{H}_n(\alpha)$ are the Hermite polynomials in the complex variable $\alpha$. Therefore, the coherent states \textcolor{beamer@PRD}{(}\ref{coherent}\textcolor{beamer@PRD}{)} appear to be a special solution of squeezed states \textcolor{beamer@PRD}{(}\ref{squeezed}\textcolor{beamer@PRD}{)} for $\gamma=0$. A direct generalisation of the above formalism is carried out by the replacement of boson creation and annihilation operators $a,a^\dagger$ with the generalised ladder operators $A, A^\dagger$, such that
\begin{alignat}{1}
A &= af(a^\dagger a)= f(a^\dagger a+1)a, \\
A^\dagger &= f(a^\dagger a)a^\dagger= a^\dagger f(a^\dagger a+1),
\end{alignat}
with $f(a^\dagger a)$ being an operator valued function of the number operator \cite{Alvarez_Hussin,Angelova_Hertz_Hussin}. For the choice of $f(a^\dagger a)=1$, the generalised ladder operators $A,A^\dagger$ fall into the canonical creation and annihilation operators $a,a^\dagger$. Note that, the operator valued function $f(a^\dagger a)$ can be associated with the eigenvalues $e_n$ of a system composed of the ladder operators $A$ and $A^\dagger$, as follows
\begin{equation}\label{similarity}
A^\dagger A=f(a^\dagger a)a^\dagger af(a^\dagger a)=f^2(a^\dagger a)a^\dagger a\sim f^2(n)n=e_n,
\end{equation} 
which holds in general for the function $f(n)$. The appearance of additional constant terms in the eigenvalues can be realised by rescaling the composite system of $A$ and $A^\dagger$ correspondingly. In the generalised framework, the coherent states \textcolor{beamer@PRD}{(}\ref{coherent}\textcolor{beamer@PRD}{)} are modified according to the eigenvalue definition, $A\vert z,f\rangle=z\vert z,f\rangle$, to
\begin{eqnarray}\label{GenCoherent}
\vert z,f\rangle &=& \frac{1}{\mathcal{N}(z,f)}\displaystyle\sum_{n=0}^{\infty}\frac{z^n}{\sqrt{n!}f(n)!}\vert n\rangle, \qquad f(0)!=1,
\end{eqnarray}
which are familiar as the nonlinear coherent states \cite{Manko_Marmo_Solimeno_Zaccaria,Filho_Vogel,Sivakumar}, with the nonlinearity arising from the function $f(n)$. The normalisation constant $\mathcal{N}(z,f)$ can be computed by the requirement $\langle z,f\vert z,f\rangle =1$. For more informations regarding the generalisation of the coherent states, we refer the readers to \cite{Dodonov,Perelomov_Book}. By following \textcolor{beamer@PRD}{(}\ref{similarity}\textcolor{beamer@PRD}{)}, also it is possible to express the coherent states \textcolor{beamer@PRD}{(}\ref{GenCoherent}\textcolor{beamer@PRD}{)} in terms of the eigenvalues of the corresponding systems as given below \cite{Gazeau_Klauder,Dey_Hussin}
\begin{eqnarray}\label{GenCoh}
\vert z,f\rangle &=& \frac{1}{\mathcal{N}(z,f)}\displaystyle\sum_{n=0}^{\infty}\frac{z^n}{\sqrt{e_n!}}\vert n\rangle, \qquad e_0=0.
\end{eqnarray}
Construction of squeezed states in the generalised formalism is more involved. Indeed, one requires to expand the squeezed states $\vert z,\gamma, f\rangle$ in terms of Fock states
\begin{alignat}{1}\label{GenSqueezed}
\vert z,\gamma, f\rangle &= \frac{1}{\mathcal{N}(z,\gamma,f)}\displaystyle\sum_{n=0}^{\infty}\frac{\mathcal{I}(z,\gamma,n)}{\sqrt{n!}f(n)!}\vert n\rangle =\frac{1}{\mathcal{N}(z,\gamma,f)}\displaystyle\sum_{n=0}^{\infty}\frac{\mathcal{I}(z,\gamma,n)}{\sqrt{e_n!}}\vert n\rangle,
\end{alignat}
followed by the substitution into the corresponding eigenvalue equation, $(A+\gamma A^\dagger)\vert z,\gamma, f\rangle = z \vert z,\gamma, f\rangle$, to yield a three term recurrence relation \cite{Angelova_Hertz_Hussin,Dey_Hussin}
\begin{equation}\label{recurrence}
\mathcal{I}(z,\gamma,n+1)-z~\mathcal{I}(z,\gamma,n)+\gamma n f^2(n)\mathcal{I}(z,\gamma,n-1)=0,
\end{equation} 
with $\mathcal{I}(z,\gamma,0)=1$ and $\mathcal{I}(z,\gamma,1)=z$. It is worthwhile to mention that due to the inadequacy of general formalism in the literature, one requires to solve the recurrence relation \textcolor{beamer@PRD}{(}\ref{recurrence}\textcolor{beamer@PRD}{)}  each time to obtain the squeezed states for the model corresponding to the particular value of $f(n)$. To this end, we would like to mention that the nonlinear coherent \textcolor{beamer@PRD}{(}\ref{GenCoherent}\textcolor{beamer@PRD}{)} and squeezed states \textcolor{beamer@PRD}{(}\ref{GenSqueezed}\textcolor{beamer@PRD}{)} are very interesting areas of research in recent days. Let us mention, for example, oscillating motion of a particle in a quadratic potential is described by the coherent states. For a non-quadratic confining potential, the nonlinear coherent state could be a good model for such motion states. Also, the nonlinear coherent state could be an adapted model for studying several physical systems with a non-harmonic potential; such as, the Bose-Einstein condensate \cite{Greiner_Etal}, Hall effects \cite{Ezawa_Book}, nonlinear field theories \cite{Ashtekar_Etal}, etc. Experimentally, the nonlinearity can be realised by the optical pumping (with a laser) of nonlinear mediums; such as, the Kerr medium inside a cavity \cite{Ourjoumtsev_etal}. The emitted light from such nonlinear mediums can be modelled as a nonlinear coherent light, which are nonclassical in nature \cite{Ourjoumtsev_etal,Baas_Karr_Eleuch_Giacobino}. The existence of squeezed states in nonlinear mediums have also been found experimentally; see, for example \cite{Huang_Etal}.
\end{section} 
\begin{section}{Nonclassicality via entanglement}\label{sec3}
The modern quantum optics is undergoing through an extensive research of detecting various nonclassical states and analysing the qualitative behaviour among them. A well known fact is that in terms of nonclassicality the coherent states define a boundary, where the nonclassical effects are absent.  There exist several approaches of examining the nonclassicality of states, for instance, by analysing quadrature and photon number squeezing \cite{Walls,Loudon_Knight} or higher order squeezing \cite{Hong_Mandel}, by examining the negativity of the Wigner function \cite{Wigner}, by testing the separability with the Peres-Horodecki criterion \cite{Duan,Simon,Agarwal_Biswas} etc. Here, we would like to implement a well established protocol, i.e. the quantum beam splitter \cite{Campos_Saleh_Teich,Kim_Son_Buzek_Knight,Dey_Hussin}. The output states of a beam splitter are realised by acting a unitary operator $\mathcal{B}$ on the input states:
\begin{equation}
\vert\text{out}\rangle =\mathcal{B}\vert\text{in}\rangle =e^{\frac{\theta}{2}(a^\dagger b e^{i\phi}-ab^\dagger e^{-i\phi})}\vert\text{in}\rangle , \qquad \vert\text{in}\rangle =\vert\psi_1\rangle\otimes\vert\psi_2\rangle ,
\end{equation}
with $a,a^\dagger$ and $b,b^\dagger$ being the sets of canonical ladder operators operating on the input fields $\vert\psi_1\rangle$ and $\vert\psi_2\rangle$, respectively. $\phi$ is the phase difference between the reflected and transmitted fields and $\theta\in[0,\pi]$ denotes the angle of the beam splitter. For more details on the device we refer the readers to \cite{Campos_Saleh_Teich}. It is well known that when we pass the coherent state \textcolor{beamer@PRD}{(}\ref{coherent}\textcolor{beamer@PRD}{)} through one of the input ports and a vacuum state $\vert 0\rangle$ through the other, we obtain two coherent states at the output ports with the intensity in each of the outputs being halved. However, passing nonclassical states through the inputs create inseparable/entangled states at the output. Thus it can be utilised as a simple and efficient tool to test the nonclassicality of the input states by checking whether the output states are entangled or not. Let us briefly review the detailed procedure of computing the outputs for arbitrary inputs. Consider, for instance, the generalised squeezed states \textcolor{beamer@PRD}{(}\ref{GenSqueezed}\textcolor{beamer@PRD}{)} at one of the inputs, such that the output states become
\begin{equation}\label{output}
\vert\text{out}\rangle =\mathcal{B}(\vert z,\gamma,f\rangle\otimes \vert 0\rangle)=\frac{1}{\mathcal{N}(z,\gamma,f)}\displaystyle\sum_{n=0}^{\infty}\frac{\mathcal{I}(z,\gamma,n)}{\sqrt{n!}f(n)!}\mathcal{B}(\vert n\rangle\otimes\vert 0\rangle),
\end{equation} 
where the action of the beam splitter on the Fock state is known \cite{Kim_Son_Buzek_Knight}
\begin{equation}\label{FockInput}
\mathcal{B}\big(\vert n\rangle\otimes\vert 0\rangle\big)=\displaystyle\sum_{q=0}^{n} \begin{pmatrix}
n \\ q
\end{pmatrix}^{1/2} t^qr^{n-q}\big(\vert q\rangle\otimes\vert n-q\rangle\big),
\end{equation}
with $t=\cos(\theta/2)$ and $r=-e^{i\phi}\sin(\theta/2)$ being the transmission and reflection coefficients, respectively. By substituting Eq. \textcolor{beamer@PRD}{(}\ref{FockInput}\textcolor{beamer@PRD}{)} into \textcolor{beamer@PRD}{(}\ref{output}\textcolor{beamer@PRD}{)} and following the similar steps as in \cite{Dey_Hussin}, we obtain the linear entropy of the output states, $S=1-\text{Tr}(\rho_a^2)$, with $\rho_a^2$ being the reduced density matrix of system $a$ obtained by performing a trace over system $b$ of the density matrix $\rho_{ab}=\vert\text{out}\rangle\langle\text{out}\vert$, as follows
\begin{eqnarray}
S &=& 1-\frac{1}{\mathcal{N}^4(z,\gamma,f)} \displaystyle\sum_{q=0}^\infty\displaystyle\sum_{s=0}^\infty\displaystyle\sum_{m=0}^{\infty} \displaystyle\sum_{n=0}^{\infty} \vert t\vert^{2(q+s)}\vert r\vert^{2(m+n)} \notag \\ 
&& \times\frac{\mathcal{I}(z,\gamma,m+q)\mathcal{I}^\ast(z,\gamma,m+s)\mathcal{I}(z,\gamma,n+s)\mathcal{I}^\ast(z,\gamma,n+q)}{q!s!m!n!f(m+q)!f(m+s)!f(n+s)!f(n+q)!}~.\label{Entropy}
\end{eqnarray}
The linear entropy $S$, which varies between $0$ and $1$ is used to measure entanglement, $0$ corresponding to the case of a pure state while $1$ refers to a maximally entangled state. Note that the phase $\phi$ between the transmitted and reflected fields is not relevant for the present case as they appear with their square of their norms in \textcolor{beamer@PRD}{(}\ref{Entropy}\textcolor{beamer@PRD}{)}. We always work on $50:50$ beam splitter, i.e. with $\theta=\pi/2$. The reason is that the symmetric beam splitter produces the highest amount of entanglement among all other cases. Moreover, we perform a comparative study between different models so that the angle of the beam splitter, $\theta$, also becomes irrelevant to us.    
\end{section}   
\begin{section}{Nonclassical models}\label{sec4} 
In this section, we construct coherent states and squeezed states of several types of models associated to the operators $A,A^\dagger$ and compare their qualitative behaviour in terms of nonclassicality. To start with, let us recall some well known facts of the harmonic oscillator along with a deformed version of it.
\begin{subsection}{Linear versus quadratic spectrum}
The general form of coherent and squeezed states associated with the model corresponding to $f(n)$ are given in \textcolor{beamer@PRD}{(}\ref{GenCoh}\textcolor{beamer@PRD}{)} and \textcolor{beamer@PRD}{(}\ref{GenSqueezed}\textcolor{beamer@PRD}{)}, respectively. The harmonic oscillator is the limit of the generalised expressions for $f(n)=1$. For the case when $f(n)=\sqrt{n}$, the corresponding eigenvalues of $A^\dagger A$ are $e_n=f^2(n)n=n^2$, which we call a deformed version of the harmonic oscillator with quadratic spectrum.
\begin{figure}
\centering   \includegraphics[width=9.0cm,height=6.0cm]{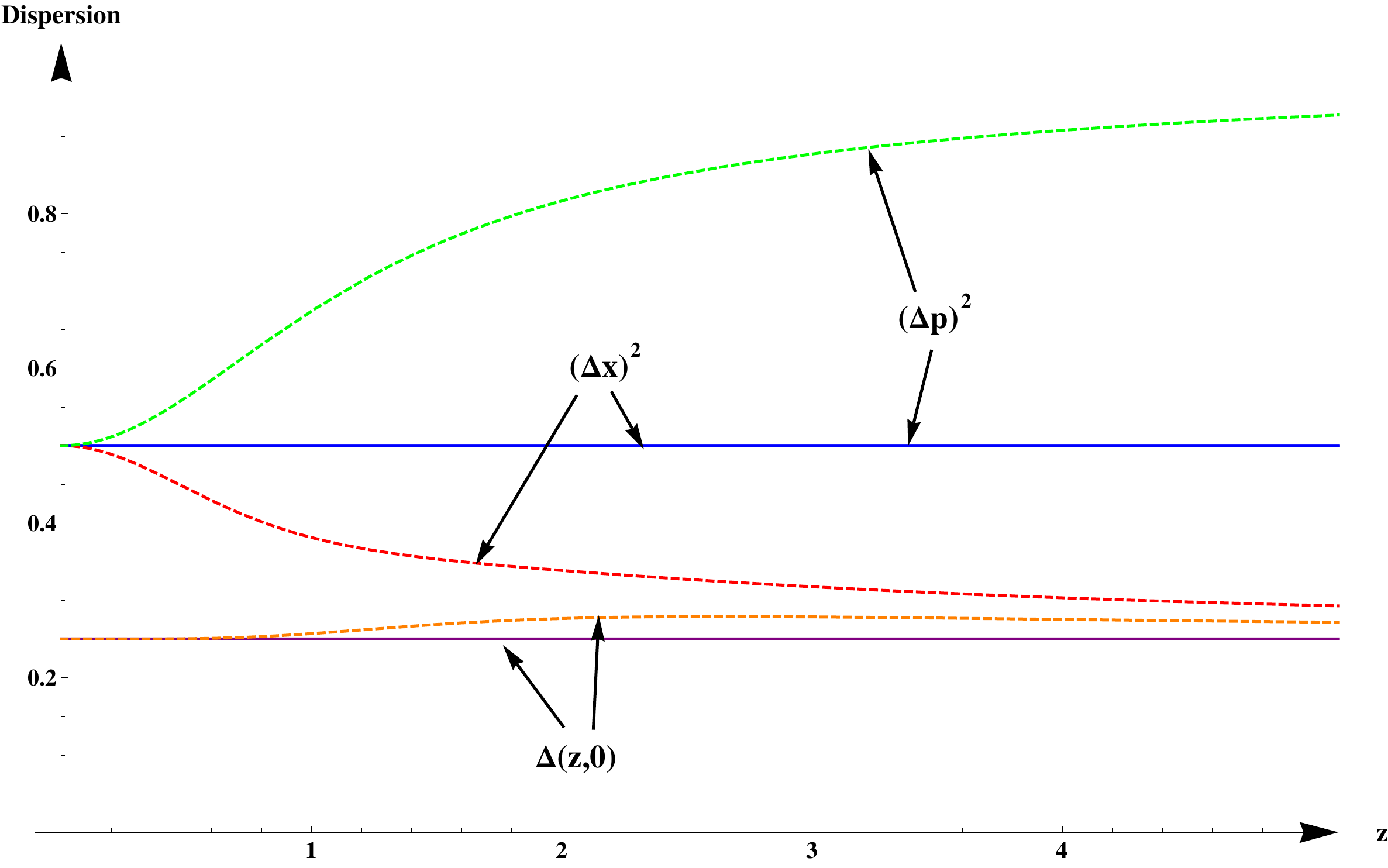}
\caption{\small{Dispersion in position $(\Delta x)^2$, momentum $(\Delta p)^2$ and product of dispersions $\Delta(z,0)=(\Delta x)^2 (\Delta p)^2$ for the usual (solid lines) and quadratic (dashed lines) coherent states of the harmonic oscillator.}}
\label{fig1}
\end{figure} 
\noindent Let us first study the squeezing properties of the above two systems in the quadrature components
\begin{equation}
x=\frac{1}{2}(a+a^\dagger), \qquad p=\frac{1}{2i}(a-a^\dagger),
\end{equation}
which are nothing but dimensionless position and momentum operators, respectively. It is well known that for the coherent states of harmonic oscillator the uncertainty relation is saturated, i.e. $\Delta x \Delta p=1/2$, with $\Delta x=\Delta p=1/\sqrt{2}$. When we consider the quadratic case, i.e. $f(n)=\sqrt{n}$, one of the quadratures $\Delta x$ becomes squeezed while the the other one $\Delta p$ is expanded correspondingly. Quadrature squeezing is a well known phenomenon, which indicates the nonclassical nature of a quantum state \cite{Walls,Loudon_Knight}. A comparative analysis of the two cases in Fig. \ref{fig1} shows that the quadratic case is nonclassical, while the harmonic oscillator resembles the classical behaviour.
  
Next we compare the density probabilities between the two cases, which are shown in Fig. \ref{fig2}. For the harmonic oscillator case, the probability density in $x$ remains uniform for all values of $z$. Whereas, in the quadratic case it becomes narrower with the increase of the value of $z$. This provides us an additional evidence of nonclassicality of the states.

Let us now move on to the squeezed states. We first solve the recurrence relation \textcolor{beamer@PRD}{(}\ref{recurrence}\textcolor{beamer@PRD}{)} for $f(n)=\sqrt{n}$, in terms of the Gauss hypergeometric function $_2F_1$ as follows
\begin{figure}
\centering   \includegraphics[width=6.5cm,height=6.0cm]{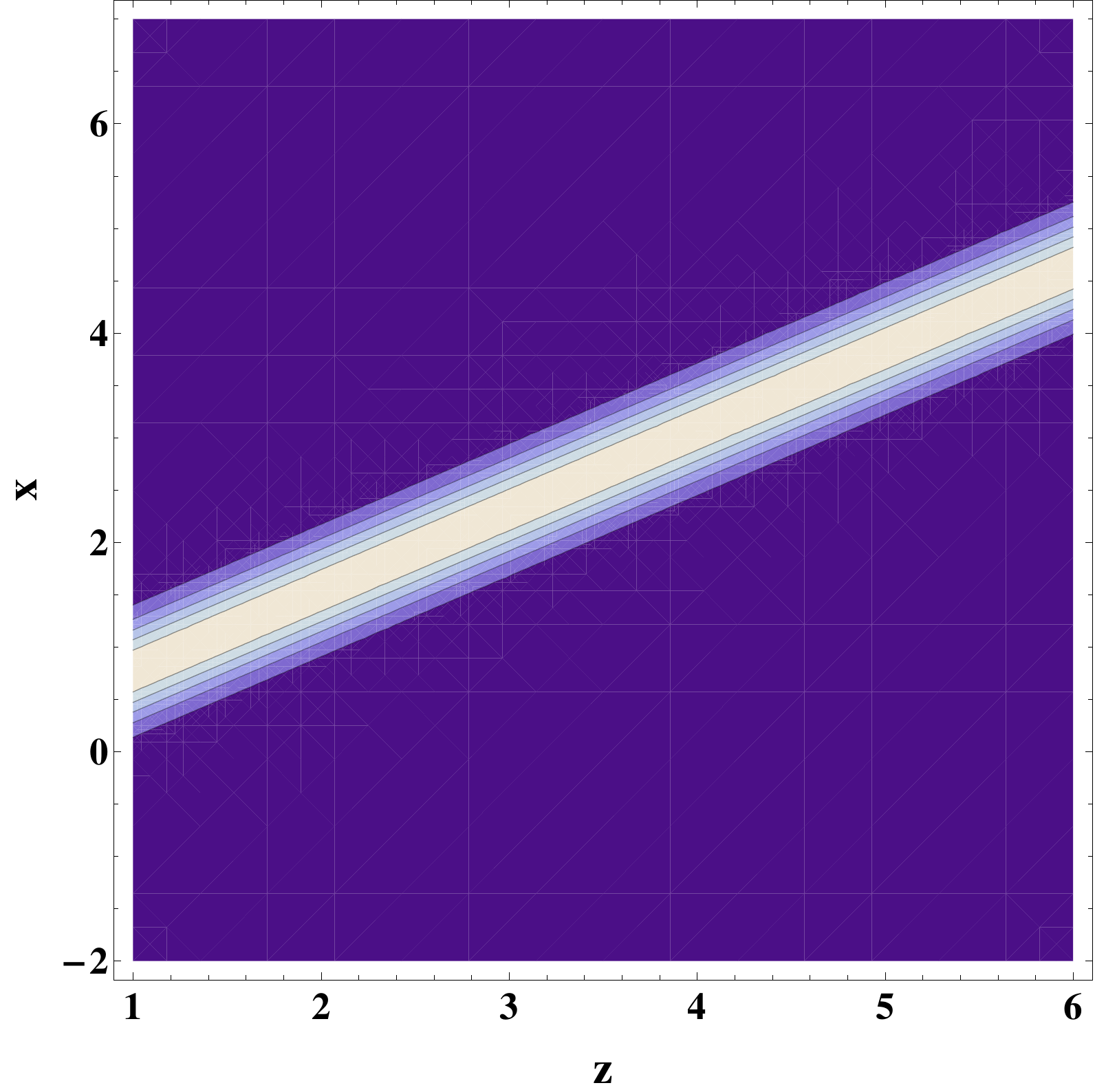}
\includegraphics[width=6.5cm,height=6.0cm]{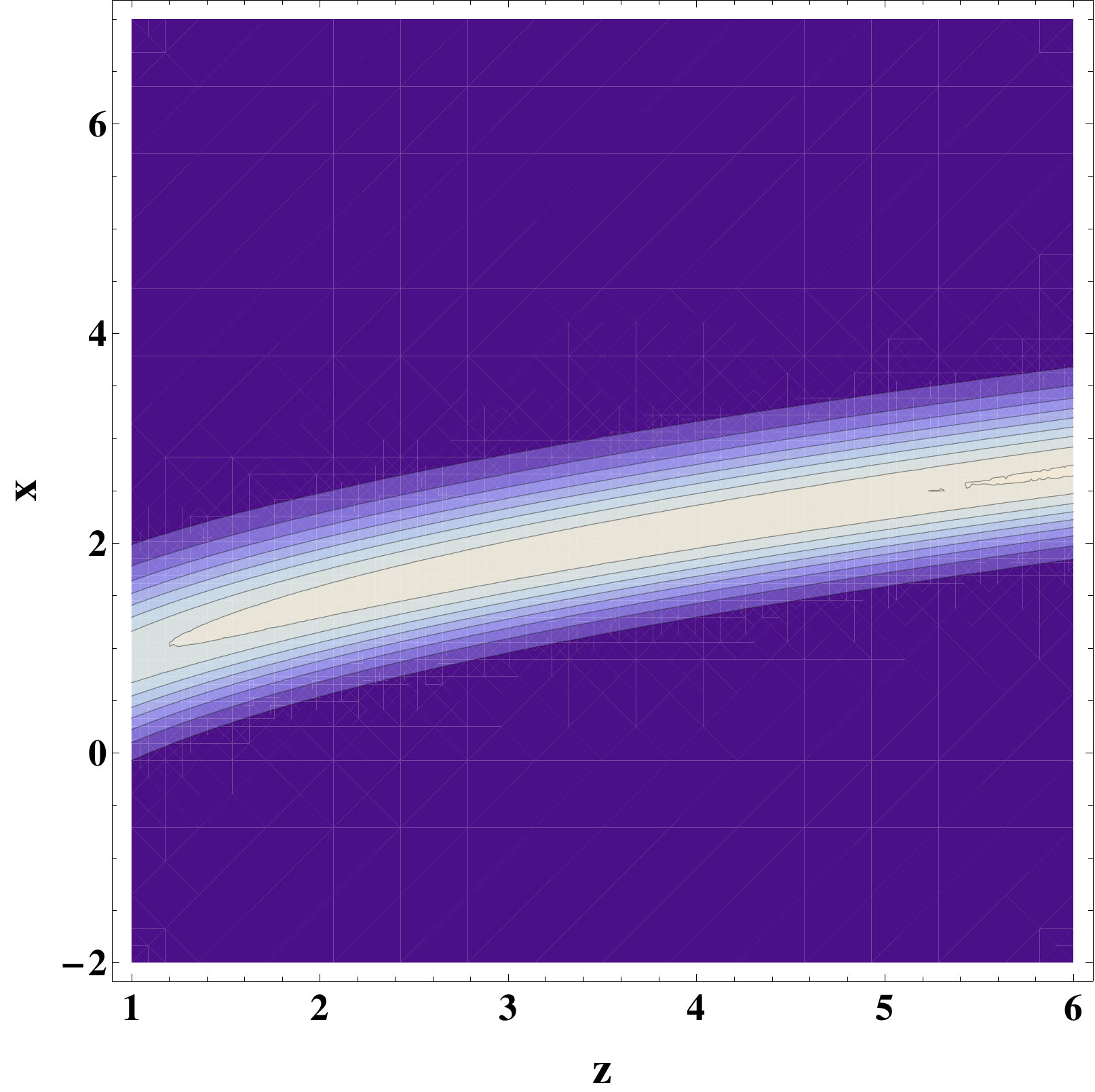}\includegraphics[width=3.0cm,height=6.0cm]{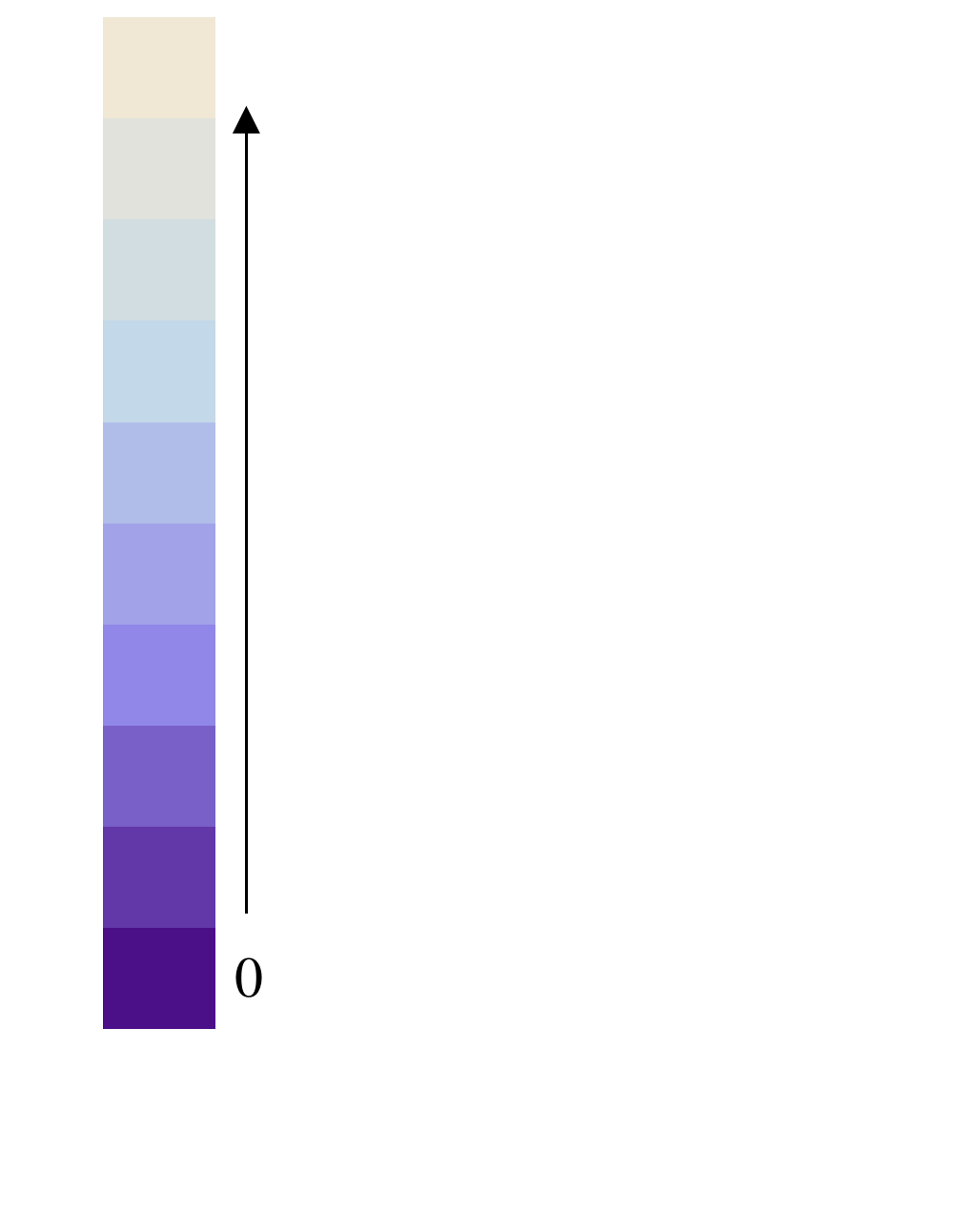}
\caption{\small{Density probabilities $\vert\psi(z,f)\vert^2$ for usual (left) versus the quadratic (right) coherent states of the harmonic oscillator.}}
\label{fig2}
\end{figure}
\begin{equation}\label{SqueezedQuadratic}
\mathcal{I}(z,\gamma,n)=i^n\gamma^{n/2}n!~_2F_1\Big[-n,\frac{1}{2}+\frac{iz}{2\sqrt{\gamma}};1;2\Big],
\end{equation}
which, when substituted in Eq. \textcolor{beamer@PRD}{(}\ref{GenSqueezed}\textcolor{beamer@PRD}{)}, we obtain the explicit form of the squeezed states for the corresponding model. We know that the squeezed states of any model are always nonclassical \cite{Walls,Loudon_Knight}. However, from the previous analysis we can claim that the squeezed states in the quadratic case are more nonclassical than those of the harmonic oscillator. To verify, we utilise the protocol discussed in Sec. \ref{sec3}. We compute the linear entropies of the output states for the input states $\vert z,\gamma\rangle\otimes\vert 0\rangle$ and $\vert z,\gamma,f\rangle\otimes\vert 0\rangle$, respectively, corresponding to the two different cases. The comparison of the outcomes are demonstrated in Fig. \ref{fig3} for different values of the squeezing parameter $\gamma$. Although for very small values of $z$, we obtain lower amount of entropy in the quadratic case than in the case of harmonic oscillator. It may happen because of the fact that the lower value of $z$ corresponds to the case when the uncertainties are almost same for both of the cases. However, we do not have a concrete argument here behind this limitation, and we leave it as an open problem. Nevertheless, our results are valid for higher values of $z$.
\begin{figure}
\centering   \includegraphics[width=9.0cm,height=6.0cm]{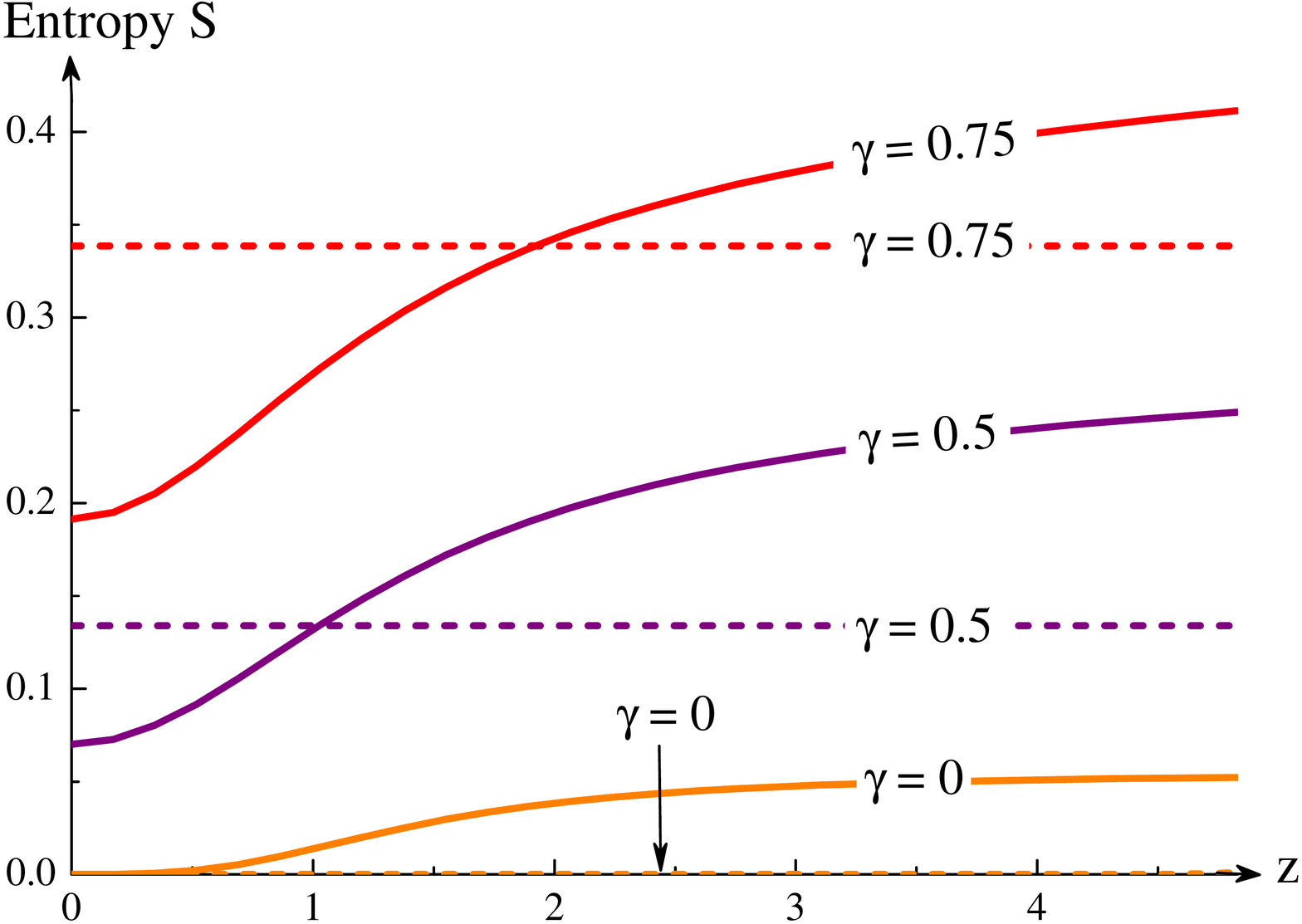}
\caption{\small{Linear entropy for squeezed states of harmonic oscillator (dashed lines) versus quadratic spectrum (solid lines).}}
\label{fig3}
\end{figure}
\end{subsection} 
\begin{subsection}{Linear plus quadratic spectrum}
In this section, we would like to explore the nonclassical behaviour of the models with eigenvalues of type $e_n=An+Bn^2$, such that $f(n)=\sqrt{A+Bn}$, where $A,B$ being some parameters, $B\neq 0$. Note that there exists various models which belong to this class of eigenvalues for different parameters, for instance, the P\"oschl-Teller model \cite{Antoine_Gazeau_Monceau_Klauder_Penson}, the harmonic oscillator in noncommutative space \cite{Dey_Fring_Khantoul}, etc. The squeezed states for the corresponding class of models are constructed by solving the recurrence relation \textcolor{beamer@PRD}{(}\ref{recurrence}\textcolor{beamer@PRD}{)} 
\begin{equation}\label{SqueezedLinQua}
\mathcal{I}(z,\gamma,n)=i^n\left(\gamma B\right)^{n/2}\left(1+\frac{A}{B}\right)^{(n)}~_2F_1\Big[-n,\frac{1}{2}+\frac{A}{2B}+\frac{iz}{2\sqrt{\gamma B}};1+\frac{A}{B};2\Big],
\end{equation}
and computing 
\begin{figure}[h]
\centering   \includegraphics[width=8.0cm,height=6.0cm]{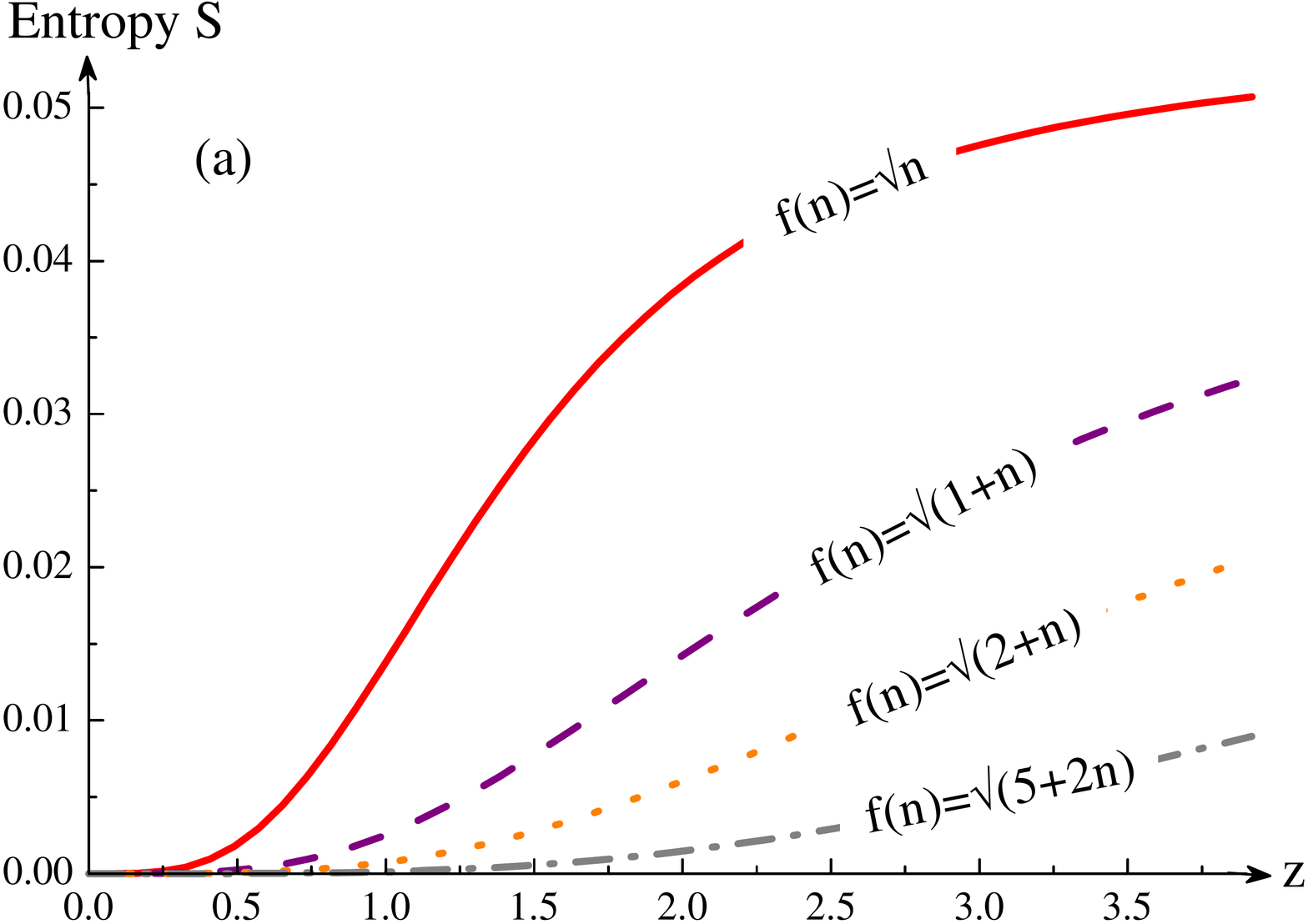}
\includegraphics[width=8.0cm,height=6.0cm]{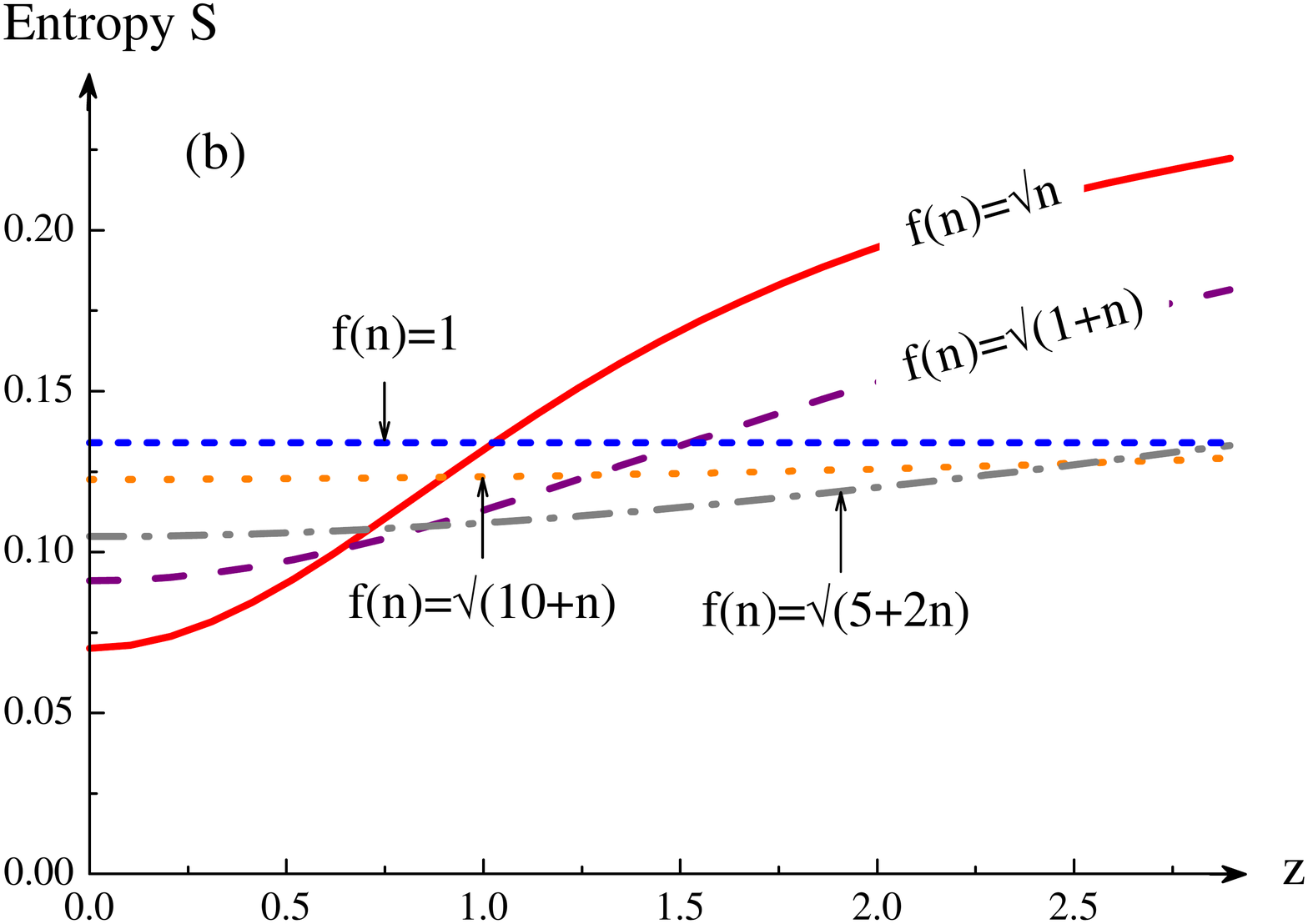}
\caption{\small{Linear entropy for (a) coherent states, (b) squeezed states for $\gamma=0.5$.}}
\label{fig4}
\end{figure}
\begin{eqnarray}
f^2(n)! &=& B^n\left(1+\frac{A}{B}\right)^{(n)},
\end{eqnarray}
where $Q^{(n)}:=\prod_{k=0}^{n-1}(Q+k)$ denotes the Pochhammer symbol with the raising factorial. We calculate the entanglement of the coherent and squeezed states of the corresponding models for different values of $A$ and $B$, as shown in Fig. \ref{fig4}\textcolor{beamer@PRD}{(a)} and Fig. \ref{fig4}\textcolor{beamer@PRD}{(b)}, respectively. Note that, the expression of linear entropy \textcolor{beamer@PRD}{(}\ref{Entropy}\textcolor{beamer@PRD}{)} contains infinite sums and, therefore, it is not an easy task to compute the entanglement. In addition, one needs to be careful when one deals with the entropy for finite number of levels. We have carefully investigated the minimum requirement of the number of levels for the convergence of the series for the corresponding values of the parameters that we have chosen in our computation. For instance, in case of Fig. \ref{fig3}, we have considered the number of levels equals to 40 and in case of Fig. \ref{fig4} it is 30. 

As indicated in the previous section, apart from the case of the harmonic oscillator we obtain finite amount of entropies for coherent states in each case corresponding to different values of the function $f(n)$ as shown in Fig. \ref{fig4}\textcolor{beamer@PRD}{(a)}. This suggests the general nonclassical nature of the nonlinear coherent states. In addition, we observe that the quadratic case, $f(n)=\sqrt{n}$, produces the highest amount of entanglement among all other cases. The similar effects are also found for the case of squeezed states as demonstrated in Fig \ref{fig4}\textcolor{beamer@PRD}{(b)}.
\end{subsection}
\end{section} 
\begin{section}{Conclusions}\label{sec5}
We have studied the nonclassical behaviour of nonlinear coherent states for several classes of models. We observed that the nonclassicality depends on the nature of the function $f(n)$. By comparing the entanglement properties of the coherent states and the squeezed states for different models corresponding to different values of $f(n)$, we showed that the model with quadratic spectrum produces the highest amount of entanglement. Thus, our analysis might help someone to find out the right models, whose degree of nonclassicality may be higher and, therefore, may produce more entanglement in comparison to other models. For further investigations, one can study few other nonclassical properties \cite{Walls,Loudon_Knight,Dey} as discussed in Sec. \ref{sec3} of our models to confirm our findings. One can also perform a similar analysis as presented here, or utilise any other powerful techniques available in the literature to verify whether the quadratic spectrum of other classes of models delivers the highest nonclassicality and entanglement, which we leave as open problems. 

\vspace{0.5cm} \noindent \textbf{\large{Acknowledgements:}} AH acknowledges the support from the Canadian NSERC research fellowship and  from the Belgian FNRS research fellowship. SD is supported by the Postdoctoral Fellowship jointly funded by the Laboratory of Mathematical Physics of the Centre de Recherches Math\'ematiques and by Prof. Syed Twareque Ali, Prof. Marco Bertola and Prof. V\'eronique Hussin. VH acknowledges the support of research grants from NSERC of Canada.
\end{section}


\end{document}